\documentclass{article}
\usepackage[version=3]{mhchem} % Formula subscripts using \ce{}
\usepackage{amssymb,amsmath}
\usepackage{bm}
\usepackage{amsfonts}
\usepackage{braket}
\usepackage{mciteplus}
\usepackage{appendix}
\usepackage{graphicx}
\usepackage{epstopdf}
\usepackage{threeparttable}
\usepackage{booktabs}
\usepackage{lscape}
\usepackage{color}
\usepackage{soul}
\usepackage{multirow}
\usepackage{setspace}
\usepackage{multirow}
\usepackage{breqn}
\usepackage{bm}
\usepackage{siunitx}
\usepackage{geometry}
\usepackage{fancyref}
\usepackage[framemethod=tikz]{mdframed}
\usepackage{authblk}
\usepackage{pdfpages}
\begin{document}

\title{Path collective variables without paths}
\author[1,2]{Dan Mendels}
\author[1,2]{GiovanniMaria Piccini}
\author[1,2]{Michele Parrinello}
\affil[1]{ Department of Chemistry and Applied Biosciences, ETH Zurich, c/o USI Campus, via Giuseppe Buffi 13, CH-6900, Lugano, Switzerland}
\affil[2]{Facolt{\'a} di Informatica, Instituto di Scienze Computationali, Universit{\'a} della Svizzera italiana (USI), Via Giuseppe Buffi 13, CH-6900, Lugano, Ticino, Switzerland}
\date{}
\maketitle

\begin{abstract}
\noindent
We introduce a method to obtain one-dimensional collective variables
for studying  rarely occurring transitions between  two metastable states separated by a high free energy barrier. No previous information, not even approximated, on the path followed during the transition is needed. The only requirement is to know the fluctuations of the system while in the two metastable states. With this information in hand we build the collective variable using a modified version of Fisher’s linear discriminant analysis.  The usefulness of this approach is tested on the metadynamics simulation of two representative systems. The first is the freezing of silver iodide into the superionic $\alpha$-phase, the second  is the study of  a classical Diels Alder reaction. The collective variable works very well in these two diverse  cases. 
\end{abstract}

\section*{Introduction}

A central problem in modern day atomistic simulations is the study of systems exhibiting complex free energy landscape in which long lived metastable states separated by kinetic bottlenecks are present. The kinetic bottlenecks impede transition between metastable states and restrict the time scale that can be explored\cite{Peters2017}.  Special purpose machines\cite{Lindorff-Larsen2011} have been constructed to push this limit.  However, in spite of remarkable progress, problems like drug unbinding or the nucleation of first order phase transitions occur on a time scale that that pose them out of the reach of direct simulations, not to mention the fact that such machines can be accessed only by a restricted number of researcher. This has motivated a vast community of modelers to develop enhanced sampling methods that allow investigating the properties of complex systems in an affordable computational time, overcoming kinetic bottlenecks and exploring different metastable states. 

 It is not practical to review here the vast literature on the subject, and we refer the reader to a recent publication for an updated review\cite{Peters2016}.  Following the pioneering work of Torrie and Valleau\cite{Torrie1977}, a large class of enhanced sampling methods rely on the identification of an appropriate set $s(\mathbf{R})$ of  collective variables (CVs) The $s(\mathbf{R})$ are appropriately chosen functions of the atomic coordinates $\mathbf{R}$. They should be able to distinguish between different metastable states and describe those modes that are difficult to sample. Their choice is crucial for a successful simulation. Many enhanced sampling methods can be described as ways of enhancing the fluctuation of the CVs  so as to favor transitions from one metastable state to another.  Two such methods are metadynamics\cite{laio_parrinello_2002,Dama2014,Barducci2008} or variationally enhanced sampling\cite{valsson_parrinello_2014} but also other methods can be similarly described\cite{valsson_2016}. 

Over the years a vast amount of experience has been gathered and a large number of CVs have been programmed and made publically available for use in connection with several enhanced sampling methods\cite{tribello_2014}.  However, whenever a new class of problems needs to be tackled, the construction of new CVs requires some effort. While this construction can be very instructive \emph{per se}\cite{Gervasio2005}, it would be useful to have a simple and effortless method to construct efficient and low dimensional CVs.   One could distinguish between two types of CVs . The first type does not assume that the final state is known a priori, and are generally used in an exploratory mode.  Typical examples are the potential energy\cite{Bonomi2010} or CVs that are used as a surrogate for the entropy\cite{Piaggi2017,Palazzesi2017}.  The second type instead assumes that the initial (A) and final states (B) are known and is used  to measure free energy profiles. Typical examples are the so-called path CVs\cite{Branduardi2007}.  

The rational behind our theory is best understood if we recall the scenario that underpins the success of finite-temperature-string method or path-CV metadynamics.  These theories are based on the   observation that in a rare event scenario the ensemble of all the reactive trajectories that go from A to B form a tube whose center lies along minimum free energy path. Finite temperature string method is based on finding the minimum free energy path while in path CV metadynamics this condition can be somewhat relaxed, as considerable empirical evidence has shown. 

 In order to understand why, we recall some of the features of path CV metadynamics. Let the FES be spanned by $N_d$ descriptors $d_i(\mathbf{R})$ that are function of the atomic coordinates $\mathbf{R}$ and are arranged to form a  vector $\mathbf{d}(\mathbf{R})$. In this free energy  space one  defines a reference path $\mathbf{d}(t)$ parametrized such that one is in  A for $t=0$ and reaches the B state when $t=1$. In ref. \cite{Branduardi2007} two CVs where defined one that measure the progression along the path ($s$) and another that measure the distance from the path  ($z$). Here only the first one is of interest and reads

\begin{equation}
\label{s_path}
s = \lim_{\lambda \rightarrow \infty} \frac{\int_0^1 t e^{-\lambda (\mathbf{d}(\mathbf{R}) -  \mathbf{d}(t))^2 } dt}{\int_0^1 e^{-\lambda (\mathbf{d}(\mathbf{R}) -  \mathbf{d}(t))^2 } dt}
\end{equation}

\begin{figure}
    \centering
    \includegraphics[width=0.5\columnwidth]{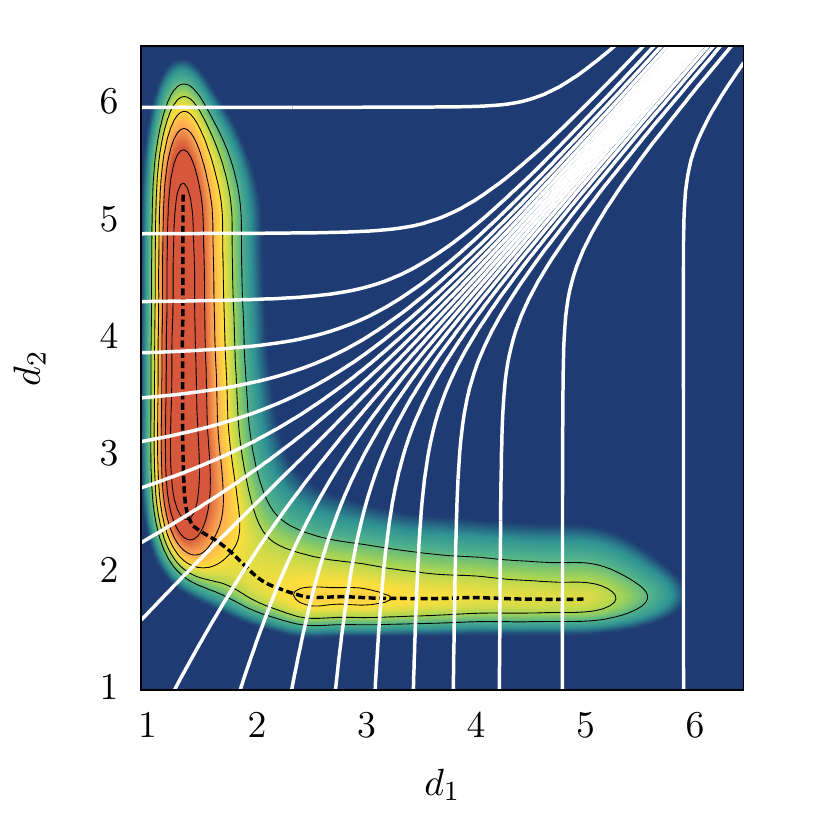}
    \caption{Free energy surface showing two basins of attraction connected by the MFEP (dashed black line). The  solid white lines are the contour isolines relative to the path variable ($s$) calculated using the MFEP as reference.}
    \label{fig:pathcv}
\end{figure}

In the first applications it was suggested to optimize $\mathbf{d}(t)$  but very soon it was realized that even reference paths  that depart within limits  from the optimal one could be successfully employed. To understand this empirical fact we note that the surfaces of constant s  partition the space into hyper surfaces  of dimension $N_d-1$ 
This is often referred to  as a foliation of the space. Close to $\mathbf{d}(t)$  the surfaces become flat and  locally orthogonal to $\mathbf{d}(t)$ (see Fig. \ref{fig:pathcv}).  In the following we shall suppose that $\mathbf{d}(t)$ threads only once the foliating surfaces.  If this were not the case, it  would signal a poor choice of descriptors, a problem that could be  possibly be solved by increasing the number of descriptors.

In order to understand the observed relative insensitivity to the choice of $\mathbf{d}(t)$ it is convenient to change  our perspective  and focus on the dual point view of  the surfaces that foliate the space rather than on  $\mathbf{d}(t)$  itself. We then imagine to start a simulation from a position not too distant from the minimum free energy path and then let the system evolve under the condition that s is kept constant. The spontaneous evolution of the system will bring it close to the minimum of the intersection between the fixed s surface and the underlining FES.   Thus a bias based on an imperfect $\mathbf{d}(t)$    still drags the system along the minimum free energy path.  Of course, the more orthogonal the foliating   surfaces are to the path direction, the more efficient the simulation. It is this self-focusing effect that we will count on in the following.

Now the stage is set to describe our approach.  If we look at the problem from the more forgiving foliation point of view, it is then worthwhile to consider the simplest possible foliation, namely the one in which the space is partitioned into planar surfaces of dimension $N_d-1$ is given.  
Any such foliation is defined, once the direction orthogonal to the planes. Let this direction be determined by the vector $\mathbf{W}$. As argued above we have some leeway in the choice of $\mathbf{W}$.  Of course the transition state region is the one requiring the most attention in this choice.  The ideal choice would to choose $\mathbf{W}$ as to be parallel to the FES discriminant surface that, passing for the transition state, distinguish between basin A and basin B.  Since calculating the FES is the very object of our method this is not possible and even if we had the FES in all its full glory calculating the discriminant would be somewhat challenging.  

Here we take an approach that has been many times been followed in quantum chemistry namely we try to infer properties of transition state from the fluctuations of the systems while in the two metastable states. The most famous example of such an approach is the celebrated Marcus theory of electron transfer. A saving grace that we do not need to know the precise position of the discriminant but only its direction.  To help us in this endeavor we call in the help of artificial intelligence methods and use linear discriminant analysis (LDA) to identify the optimal direction in which to foliate the space for the purpose of driving the reaction.  This type of analysis was introduced to classify a set of data into two classes, and recently used to distinguish between biomolecules conformations\cite{Sakuraba2016}. Not surprisingly, since here the purpose is different, we had to modify for optimal performance   the original LDA applied to distinguish between different protein formulation and introduce a variant that we call harmonic linear discriminant analysis (HLDA).  We exemplify the usefulness of HLDA on a crystal phase transformation of the  ionic compound AgI  and to a prototypical Diels Alder reaction, before finishing the paper with some discussion.

\section*{Methods}

\subsection*{Linear discriminant analysis for metastable states}

As discussed above we assume that the system is well described once the free energy surface $F(\mathbf{d})$ is a function of the $N_d$ descriptors. On this surface lie the metastable states A and B. While trapped in these two states the descriptors will have and expectation value $\boldsymbol{\mu}_{A,B}$ and a multivariate variance $\boldsymbol{\Sigma}_{A,B}$. These quantities can be evaluated in short unbiased runs. The data generated in such runs will be separated in the $N_d$ dimensional space of the descriptors. Our first goal is to find a one dimensional projection of these two sets of data along which they still do not overlap. If we take a generic projection $x=\mathbf{W}^T \mathbf{d}$, where $\mathbf{W}$ is a vector in the $N_d$ space there is no guarantee that they will not overlap.
The linear discriminant analysis aims at finding the direction $\mathbf{W}$ along which the projected data are at best separated.   To do this one needs a measure of their degree of separation. Following Fisher this is given by the ratio between the so called between class $S_b$ and the within class $S_w$ scatter matrices.  The former is measured by the square of the distance  between the projected means

\begin{equation}
\label{mean_transf}
\boldsymbol{\tilde{\mu}}_A-\boldsymbol{\tilde{\mu}}_B = \mathbf{W}^T \left( \boldsymbol{\mu}_A - \boldsymbol{\mu}_B \right)
\end{equation}

\noindent
and  can be written as

\begin{equation}
\label{mean_transf_mat}
\mathbf{W}^T \mathbf{S}_b \mathbf{W}
\end{equation}

\noindent
where

\begin{equation}
\label{between_class}
\mathbf{S}_b = \left( \boldsymbol{\mu}_A - \boldsymbol{\mu}_B \right)\left( \boldsymbol{\mu}_A - \boldsymbol{\mu}_B \right)^T
\end{equation}

\noindent
The within spread is  estimated from  the sum of the two spreads  leading to the expression 

\begin{equation}
\label{cov_transf_mat}
\mathbf{W}^T \mathbf{S}_w \mathbf{W}
\end{equation}

\noindent
with 

\begin{equation}
\label{within_class}
 \mathbf{S}_w = \boldsymbol{\Sigma}_A + \boldsymbol{\Sigma}_B.
\end{equation}

\noindent
The Fisher’s object function then reads like a Rayleigh ratio
 
\begin{equation}
\label{fisher_ration}
\mathcal{J(\mathbf{W})} = \frac{\mathbf{W}^T \mathbf{S}_b \mathbf{W}}{\mathbf{W}^T \mathbf{S}_w \mathbf{W}}
\end{equation}

\noindent
that is maximized by

\begin{equation}
\label{maximizer}
\mathbf{W}^* = \mathbf{S}_w^{-1} \left( \boldsymbol{\mu}_A - \boldsymbol{\mu}_B \right).
\end{equation}

\noindent
Following up on the introductory discussion, this would suggest that a useful one dimensional CV is

\begin{equation}
\label{maximizer_artihmetic}
s_{LDA}(\mathbf{R}) = \left( \boldsymbol{\mu}_A - \boldsymbol{\mu}_B \right)^T \left( \boldsymbol{\Sigma}_A + \boldsymbol{\Sigma}_B \right)^{-1}   \mathbf{d}(\mathbf{R}).
\end{equation}

As we shall see below the performance of such CV proved to be mediocre. One possible explanation could be that when taking the arithmetic averages of the covariances , it is the larger one that carries more weight. Froma a data analysis point of view this seems a bit counterintuitive since the data with smaller variance are better defined. Also from the rare event point it would seem more appropriate to give more weight to the state with the smaller fluctuations, that is more difficult to get into and escape from. For these reasons we propose a different measure of the scatter and rather the using of the arithmetic average we base the measure of the within scatter matrix on the harmonic average as follows:

\begin{equation}
\label{harmonic_mean}
 \mathbf{S}_w = \frac{1}{\frac{1}{\boldsymbol{\Sigma}_A} + \frac{1}{\boldsymbol{\Sigma}_B}}.
\end{equation}

\noindent
This lead to a different expression for the collective variable that reads:

\begin{equation}
\label{maximizer_harm}
s_{HLDA}(\mathbf{R}) = \left( \boldsymbol{\mu}_A - \boldsymbol{\mu}_B \right)^T \left( \frac{1}{\boldsymbol{\Sigma}_A} + \frac{1}{\boldsymbol{\Sigma}_B} \right)   \mathbf{d}(\mathbf{R}).
\end{equation}

From a machine learning point of view the more compact states are better defined and thus have a larger weight in determining the discriminants.
In our experience this second choice has proven to be far superior.

\section*{Results}

\subsection*{Liquid/Superionic Phase Transition in AgI}

The first example we have chosen to apply our method to is AgI. In particular we study the transition from the liquid to the $\alpha$-phase.  In this interesting transition the iodine anions order to form a BCC lattice while the silver cations remain highly mobile. This partially ordered phase is  an example of a  superionic phase.

Recently we have studied this compound using metadynamics\cite{Mendels2018} and found that a free energy surface can be drawn using enthalpy $s_H$ and a surrogate for the entropy $s_S$ as Collective Variables (CVs).  The ability of representing this complex transition with only two CVs offers the possibility of demonstrating how our prescription for constructing a one-dimensional CV works. 

\begin{figure}
	\centering
    \includegraphics[width=1.\columnwidth]{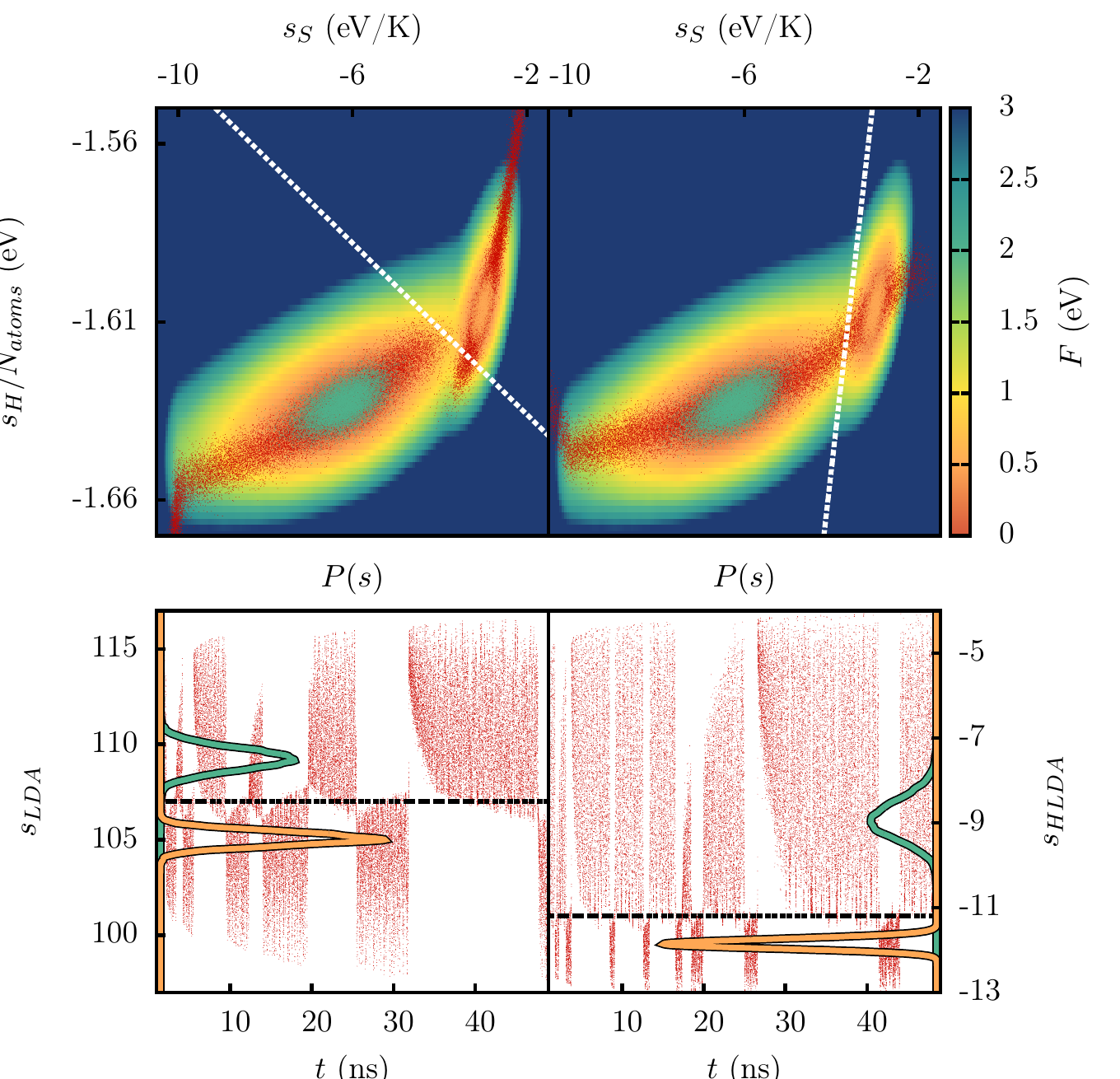}   
    \caption{Upper panel: free-energy surface along entropy, $s_S$, and enthalpy, $s_H$, collective variables. The green and orange dots represent the points sampled in the unbiased basins, the red dots are the points sampled using metadynamics in with a standard (arithmetic average) LDA CV, left panel, and HLDS, right panel. The dashed lines are the distribution discriminants to whom the CV is normal
    Lower panel: time evolution of the arithmetic LDA CV, left panel, and of the HLDA CV, right panel. The green and orange lines represent the projection of the unbiased distribution on the discriminant plane.}
    \label{fig:AgI_plot}
\end{figure}

The fluctuations around these two minima are, to a good approximation, Gaussians and are be described by their fluctuation matrices $\mathbf{\Sigma}_A$ and $\mathbf{\Sigma}_B$.  In Fig. \ref{fig:AgI_plot} the points sampled during these two unbiased runs are shown and are clearly separated. 

We can now construct two different one dimensional CVs, one built using LDA $s_{LDA}$ (Eq. \ref{maximizer_artihmetic}) and another  using  HLDA  $s_{HLDA}$  (Eq. \ref{maximizer_harm}). In Fig. \ref{fig:AgI_plot} we contrast  the orientations of the line orthogonal to the direction defined by  W.  They appear to be rather different (the computational details of the calculations can be found in the supporting information). In the transition region between A and B  this different orientation has significant influence on the quality of the CV . In the case of $s_{HLDA}$ the point sampled bunch nicely to form a tube whose center closely follow the minimum free energy path. In the $s_{LDA}$ case instead the points in the transition path branch out to follow different pathways. This a reflection of a somewhat hysteretic behavior, as can be seen in the lower panels of Fig. \ref{fig:AgI_plot}, where it is also shown the that points generated in the first  two unbiased runs are better separated in the one dimensional projection along $s_{HLDA}$.

\subsection*{Diels-Alder Reaction: [4+2] Cycloaddition of 1,3-Butadiene and Ethene}

As an example of application to chemical reactions we study a classical Diels-Alder process, namely the [4+2] Cycloaddition of 1,3-butadiene and ethene.
The essence of a chemical reaction can be summarized as a combination of simultaneous bonds breaking and formation.
This is accompanied by a time crucial conformational reorganization\cite{Piccini2017} and a local change of distances.
Such a bond reorganization can be a dramatic event involving breaking and/or formation of $\sigma$-type bonds but can also be accompanied by local electronic rearrangements resulting in a non-local strengthening or weakening of other bonds like the $\pi$ to $\sigma$ conversion.

\begin{figure}
    \centering
    \includegraphics[width=0.5\columnwidth]{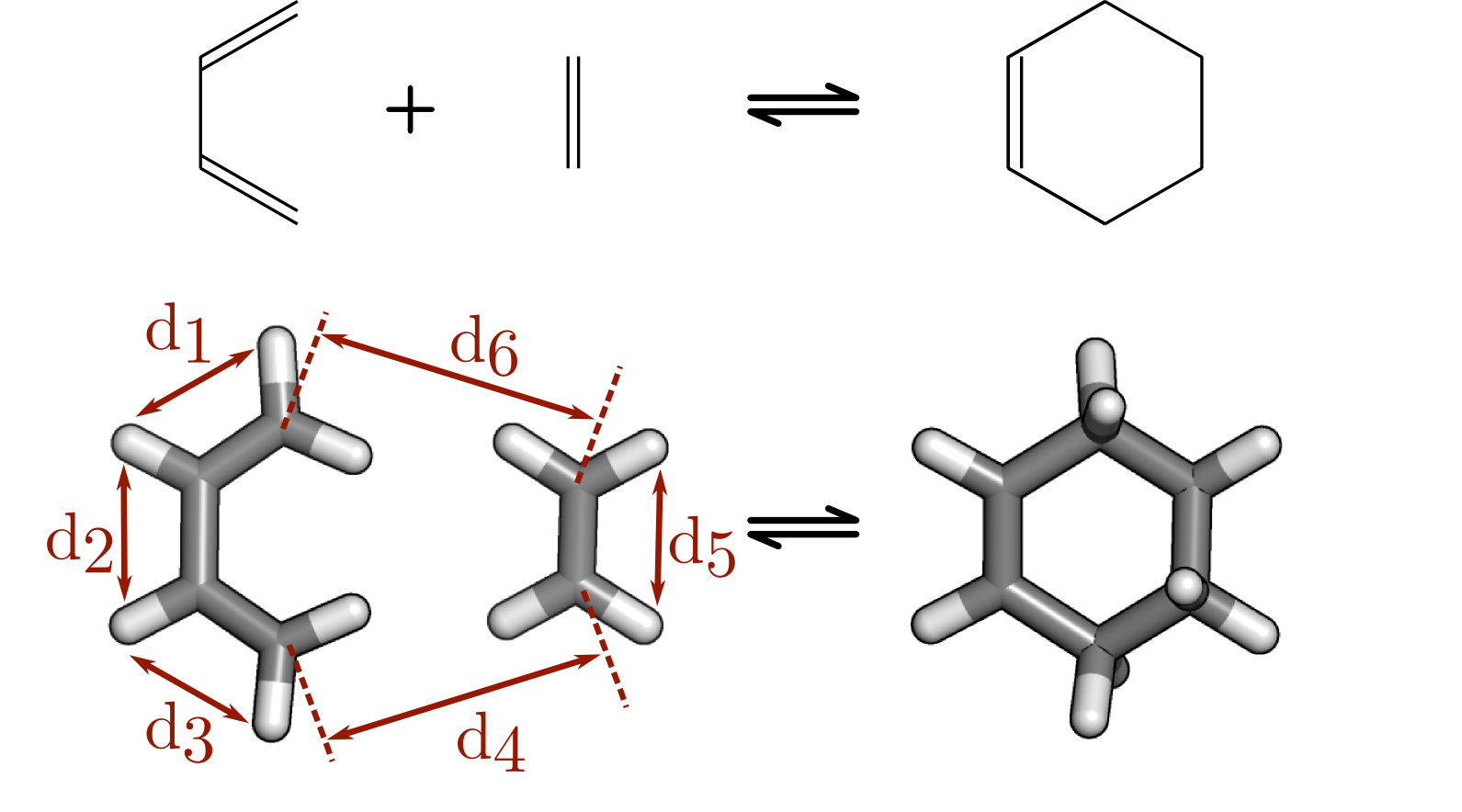}
    \caption{Reaction scheme of the [4+2] cycloaddition of 1,3-butadiene and ethene. In the stick model in the lower part of the figure the the distances used as descriptors of the process are reported.}
    \label{fig:reaction}
\end{figure}

The chemical scheme for a classical Diels-Alder reaction is reported in Fig. \ref{fig:reaction}.
The peculiarity of a Diels-Alder reaction lays in the fact that the formation of the two $\sigma$ bonds in the product state implies that three $\pi$ bonds of the reactant state become two $\sigma$ bonds while a $\sigma$ bond becomes a $\pi$ bond.
Therefore, although it is clear from Fig. \ref{fig:reaction} that distances $d_4$ and $d_6$ will play a major role in the reaction progression as they govern the approach of the two molecules to each other, the concerted elongation-contraction of distances $d_1$, $d_2$,$d_3$, and $d_5$ will play a significant role.

Thus, we used these six parameters as a basis for our discriminant analysis performing two unbiased runs of 15 and 1 ps for the reactant and product states respectively.
The fact that the reactant state needs longer simulation times for the unbiased run is due to its larger conformational freedom as the two partners in the gas-phase can explore more space than in the product states.
Since, in principle, in the gas-phase the reactants molecules can be uniformly distributed all over the volume, we applied a harmonic restraints on the $d_4$ and $d_6$ distances for values larger that 3 \si{\angstrom}. This allows the reactants to explore a significant part of the conformational space.

Having these two distributions for the six distances considered we applied HLDA to them.
In our hands the result is qualitatively intuitive.
The coefficients of the linear combination are reported in Table \ref{table:DA_coeffs} show that the largest contributions are related to the $d_4$ and $d_6$ distances that have same sign and approximately same amplitude indicating that in the concerted mechanism the two partners have to come close simultaneously.
Moreover, they show that while for instance $d_4$ and $d_6$ are shrinking the $d_2$ distance is shrinking as well as it goes from a $\sigma$-type bonding to a $\pi$-type one.
At the same time, all the bonds that go from a shorter $\pi$-type to a longer $\sigma$-type, namely distances $d_1$, $d_3$, and $d_5$, have an opposite sign reflecting the fact that they must elongate while the other three are shrinking and viceversa.

\begin{table}[htbp]
\caption{HLDA coefficients for the reaction collective variable.}
\centering
\begin{tabular}{rcccccc}
\toprule
              & $d_1$ & $d_2$ & $d_3$ & $d_4$ & $d_5$ & $d_6$  \\ 
\midrule
$\mathbf{W}^*$  & 0.18 &-0.46 & 0.13  & -0.60 & 0.05  & -0.61 \\
\bottomrule
\end{tabular}
\label{table:DA_coeffs}
\end{table}

\begin{figure}
	\centering
    \includegraphics[width=0.5\columnwidth]{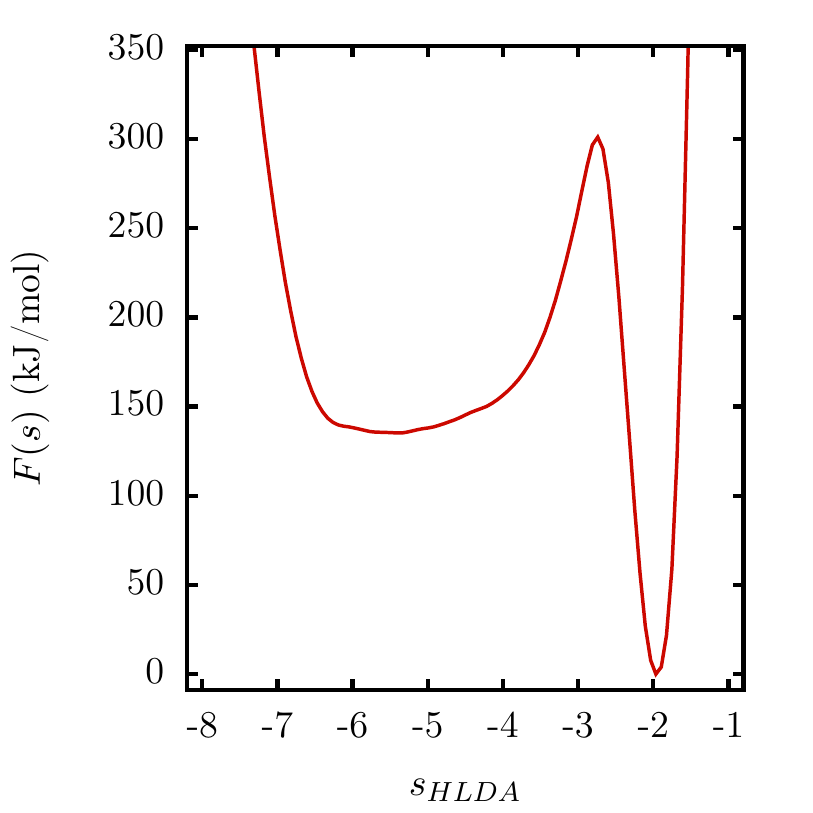}   
    \caption{Free-energy profile for the [4+2] cycloaddition of 1,3-butadiene and ethene along the HLDA CV.}
    \label{fig:DA_plot}
\end{figure}

\begin{figure}
    \centering
    \includegraphics[width=0.5\columnwidth]{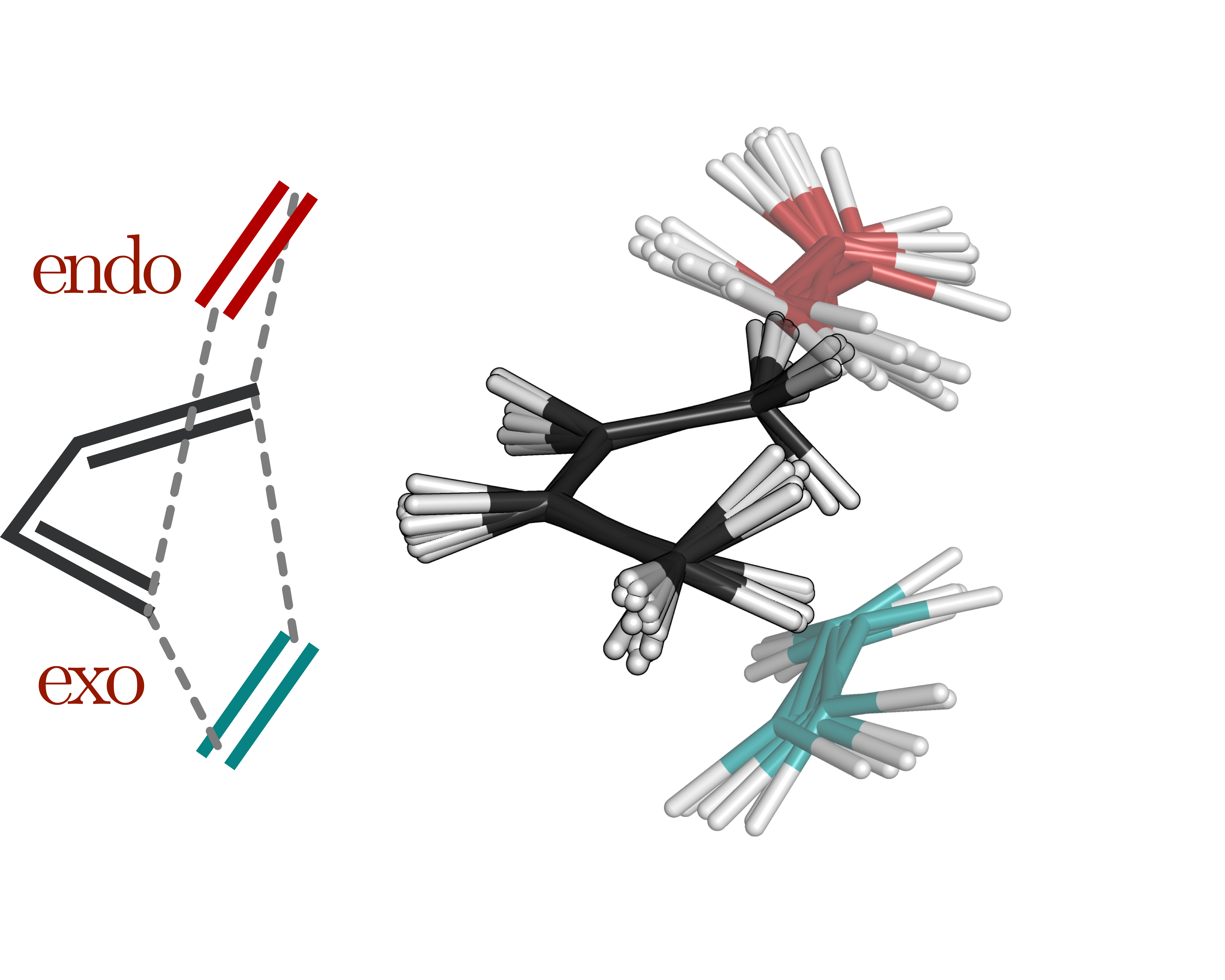}
    \caption{Overlap of an ensemble of configurations extracted from the barrier top region ($s(\mathbf{R}) \approx -2.8$) resembling the typical text book example of a Diels-Alder transition structure. Here the potentially chiral stereospecific configurations \emph{endo} and \emph{exo} are depicted in magenta and cyan.}
    \label{fig:ts_da}
\end{figure}

To demonstrate the power of such a method we report in Fig. \ref{fig:DA_plot} the Free-energy profile along $s_{HLDA}$  (the computational details of the calculations can be found in the supporting information).
The HLDA CV besides being rather efficient clearly brings out the chemistry of the problem with a wide entropic basin and a narrow and deep enthalpic state. 
Furthermore, the configurations extracted from those that are in the apparent transition state do correspond to those that are described in standard text books (see Fig. \ref{fig:ts_da}).
In the final state two possible configurations are possible, namely \emph{endo} and \emph{exo}.
In our case they are symmetry related by reflection.
However, if some specific asymmetry of the reactants would be present this may lead to the well known endo-exo chirality.
Such a procedure can be applied to more complex cases where the bonding transformation cannot be clearly attributed by means of simple chemical intuition.
Moreover, as the discrimination of the states is so well defined the method can be applied to understand more deeply the nature of the transition state that is key in the understanding of many chemical properties.

\section*{Conclusions}

In this contribution we present a simple way of obtaining efficient collective variables, to be used in collective variables based enhanced sampling methods.  The only information needed are a set of descriptors that identify the metastable states of the system and the fluctuations of these descriptors while in the basins. We use a modified version of a method often used in statistics to classify data into different classes, namely Fisher’s classical LDA. In the space of descriptors, the direction orthogonal to the hyper plane that best separate the sets of data is the sought-after collective variable.  Arguments are given why this procedure should be useful.
The validity of the arguments are borne out by two very successful calculation to two typical problems in the field of rare events nucleation and chemical reactions. The CV obtained by using our HLDA approach perform very satisfactorily and lead to highly meaningful results. The reasons for this success will need to be investigated further since they interrogate us on the nature of rare transitions, in particular for what concerns chemical reactions. 

Much practical work needs to be done also on the practical side LDA although modified to the present HLDA version is probably the simplest approach to statistical classification of data.  More sophisticated methods are possible and might lead to further improvement, also the extension of LDA to multiple class cases  might come handy in all those case in which several transitions are possible.

\newpage
\bibliographystyle{ieeetr}
\bibliography{library}
%\printbibliography

\newpage


\section*{Supplementary material}

# Supporting Information for: Path collective variables without path

GiovanniMaria Piccini[1,2], Dan Mendels[1,2], and Michele Parrinello[1,2]

[1] Department of Chemistry and Applied Biosciences, ETH Zurich, c/o USI Campus, via Giuseppe Buffi 13, CH-6900, Lugano, Switzerland
[2]Facoltá di Informatica, Instituto di Scienze Computationali, Universitá della Svizzera italiana (USI), Via Giuseppe Buffi 13, CH-6900, Lugano, Ticino, Switzerland

## Computational details

### Liquid/Superionic Phase Transition in AgI

Simulations were carried out using LAMMPS [1] patched with PLUMED 2 [2] in a simulation box consisting of 250 silver ions and 250 iodine ions using the interaction potential of ref. [3]. A time-step of 1 fs was used and constant temperature of $T = 900K$ was maintained using the velocity rescaling thermostat of Bussi et al. [4] with a relaxation time of 0.1 ps. A constant pressure of 1000 bar was additionally kept using the Parrinello-Rahman method [5, 6] with a relaxation time of 1ps to which a pressure correction term was added to set the densities close to experimental value as was also used in ref. [3]. we constrained the diagonal terms of the $h$ matrix of the Parrinello-Rahman Lagrangian to be coupled and the off-diagonal ones to zero. This avoided the occurrence of inconveniently oblong simulation boxes in the liquid phase [7]. Long range electrostatic interactions were calculated with a standard Ewald summation with a real space cutoff of 10Å and a root mean squared force accuracy of $10^{-4}$. $s_S$ and $s_H$ were calculated according to their corresponding given formulas in ref. [8] where $s_S$ was calculated using a radial cutoff of $r_{max}$=12.5Å. Metadynamics simulations were run with a bias factor $\gamma = 60$, a hill deployment time interval $\tau_G = 500fs$ and initial hill height of $W = 0.05eV$. The employed hills width was $\sigma = 0.1$.

### Diels-Alder Reaction: [4+2] Cycloaddition of 1,3-Butadiene and Ethene

Simulations were carried out using CP2K [9] patched with PLUMED 2 [2] in a non-periodic simulation box of $15 \times 15 \times 15$ Å. A time-step of 0.5 fs was used and constant temperature of $T = 300K$ was maintained using the velocity rescaling thermostat of Bussi et al. [4] with a relaxation time of 0.1ps. Metadynamics simulations were run with a bias factor $\gamma = 80$, a hill deployment time interval $\tau_G = 50fs$ and initial hill height of $W = 5.0kJ/mol$. The employed hills width was $\sigma = 0.2$.

1


\end{document}